\documentclass{article}

\usepackage{amssymb}
\usepackage{natbib}
\usepackage{bm}
\usepackage{subfigure}
\usepackage{graphicx}
\usepackage[hmargin=3cm,vmargin=3.5cm]{geometry}
\usepackage{color}

\begin{document}

\title{High-Dimensional Bayesian Regularized Regression with the {\tt bayesreg} Package}
\author{Enes Makalic \and Daniel F. Schmidt}
\maketitle 

\begin{abstract}
Bayesian penalized regression techniques, such as the Bayesian lasso and the Bayesian horseshoe estimator, have recently received a significant amount of attention in the statistics literature. However, software implementing state-of-the-art Bayesian penalized regression, outside of general purpose Markov chain Monte Carlo platforms such as {\tt Stan}, is relatively rare. This paper introduces {\tt bayesreg}, a new toolbox for fitting Bayesian penalized regression models with continuous shrinkage prior densities. The toolbox features Bayesian linear regression with Gaussian or heavy-tailed error models and Bayesian logistic regression with ridge, lasso, horseshoe and horseshoe$+$ estimators. The toolbox is free, open-source and available for use with the {\tt MATLAB} and {\tt R} numerical platforms.
\end{abstract}

\def\T{{ \mathrm{\scriptscriptstyle T} }}

\section{Introduction}

Bayesian penalized regression techniques for analysis of high-dimensional data have received a significant amount of attention in the statistics literature. Recent examples include the Bayesian lasso~\citep{ParkCasella08,Hans09}, the normal-gamma estimator~\citep{GriffinBrown10}, the horseshoe and horseshoe$+$ estimators~\citep{CarvalhoPolson10}, and the generalized double Pareto estimator~\citep{ArmaganEtAl13a}, among others. These estimators use sparsity inducing prior distributions for the regression parameters and are commonly applied in the setting of big data, where most of the predictor variables are assumed to be unassociated with the outcome. Sparsity inducing priors are implemented with exchangeable Gaussian variance mixture distributions, and the corresponding Bayesian posterior inferences often directly correspond to well-known penalized regression techniques, such as the lasso~\citep{Tibshirani96,Tibshirani11} and the elastic net~\citep{ZouHastie05}. For a comprehensive catalog of Bayesian penalized regression techniques and their frequentist analogues, see \citep{PolsonScott10,PolsonScott12a}. 

There are many software tools for fitting standard penalized regression estimators, with {\tt glmnet}~\citep{FriedmanEtAl10} and {\tt ncvreg}~\citep{BrehenyHuang11} being the two most popular implementations in practice. The software toolbox {\tt glmnet} implements the lasso and elastic net penalty for generalized linear models, while {\tt ncvreg} provides an efficient algorithm for fitting MCP~\citep{Zhang10} or SCAD~\citep{FanLi01} regularization paths in linear and logistic regression models. Both of these implementations are freely available open-source packages for the R statistical platform. In contrast, software for Bayesian penalized regression, outside of general purpose Markov chain Monte Carlo (MCMC) platforms such as {\tt WinBUGS} and {\tt Stan}, is scarce (see Section~\ref{sec:discussion}). Given the excellent theoretical properties of Bayesian penalized regression methods, it would be of great benefit to the research community if a software toolbox implementing these approaches was made available. To this end, the main contribution of this manuscript is a software toolbox that implements state-of-the-art Bayesian penalized regression estimators for linear and logistic regression models. This toolbox is free, open-source and features significant computational advantages over existing toolboxes. Further, the toolbox is implemented for both the {\tt MATLAB} and the {\tt R} numerical computing platforms.  

A technical description of the Bayesian regression toolbox is now given. Formally, consider the following Bayesian regression model for data ${\bf y} = (y_1, \ldots, y_n)^\T \in \mathbb{R}^n$ given a matrix ${\bf X} = ({\bf x}_1, \ldots, {\bf x}_n)^\T\in \mathbb{R}^{n \times p}$ of $p$ predictor variables:
\begin{eqnarray}
		z_i | {\bf x}_i, \bm{\beta}, \beta_0, \omega_i^2, \sigma^2 &\sim& \mathcal{N}_n ({\bf x}_i^\T \bm{\beta} + \beta_0, \sigma^2 \omega_i^2), \label{eqn:y} \\
		\sigma^2 &\sim& \pi(\sigma^{2}) \, d\sigma^2, \label{eqn:sigma2} \\
		\omega_i^2 &\sim& \pi(\omega_i^{2}) \, d\omega_i^2, \label{eqn:omega} \\
		\beta_0 &\sim& d\beta_0, \label{eqn:beta0} \\
		\beta_j | \lambda_j^2, \tau^2, \sigma^2 &\sim& \mathcal{N}(0, \lambda_j^2 \tau^2 \sigma^2), \label{eqn:beta} \\
		\lambda^2_j  &\sim& \pi(\lambda^2_j) \, d\lambda^2_j, \\
		\tau^2 &\sim& \pi(\tau^2) \, d\tau^2, \label{eqn:tau2}
\end{eqnarray}
where $i=(1,\ldots,n)$, $j = (1, \ldots, p)$, $\beta_0 \in \mathbb{R}$ is the intercept parameter, $\bm{\beta} \in \mathbb{R}^p$ are the regression coefficients, and the variables $(z_1 , \ldots, z_n)$ are set appropriately depending on whether the data ${\bf y}$ is continuous (see Section~\ref{sec:linear:regression}) or binary (see Section~\ref{sec:logistic}). In big data problems, the sample size $n$ is often less than the number of predictors (i.e., $p \gg n$), and the predictor matrix ${\bf X}$ is not full rank. 

The hierarchy (\ref{eqn:y})--(\ref{eqn:tau2}) consists of two key groups: (i)~the model for the sampling distribution of the data (\ref{eqn:y})--(\ref{eqn:omega}) and (ii)~the prior distributions for the regression coefficients (\ref{eqn:beta})--(\ref{eqn:tau2}). Statistical models for both the data and the prior distributions are constructed from exchangeable Gaussian variance mixture distributions~\citep{AndrewsMallows74}. The probability density function of a Gaussian scale mixture random variable $Z$ can be written as
\begin{equation}
\label{eqn:gaussian:mixtures}
	\pi_{Z}(z | \mu, \sigma^2) = \int_0^{\infty} {\rm N}(z | \mu, g(\lambda) \sigma^2) \pi_{\lambda}(\lambda) \; d\lambda,
\end{equation}
where $N(\cdot | \mu, \sigma^2)$ denotes the Gaussian distribution with mean $\mu \in \mathbb{R}$ and variance $\sigma^2 > 0$, $g(\lambda)$ is a positive function of the mixing parameter $\lambda$, and $\pi_{\lambda}(\cdot)$ is a mixing density defined on $\mathbb{R}^+$. Particular choices of the mixing parameter $\lambda$ and the mixing density $g(\lambda)$ can be used to model a wide variety of non-Gaussian distributions. For example, selecting the gamma distribution ${\rm Ga}(\delta/2, \delta/2)$ as the mixing density and setting $g(\lambda) = 1/\lambda$ yields the Student $t$ density with $\delta > 0$ degrees of freedom. This decomposition may alternatively be written in the following hierarchical form:
\begin{equation}
	\label{eqn:studentt}
	Z | \mu, \sigma^2, \lambda \sim {\rm N}(\mu, \sigma^2/\lambda), \quad \quad \lambda | \delta \sim {\rm Ga}(\delta/2, \delta/2).
\end{equation}
There are many distributions that admit the scale mixture of Gaussians representation; for example, the logistic density~\citep{PolsonEtAl13}, the Laplace distribution~\citep{AndrewsMallows74}, and the class of $z$ distributions~\citep{BarndorffNielsenEtAl82}.

In the case of the data model (\ref{eqn:y})--(\ref{eqn:omega}), the scale parameter $(\sigma^2 > 0)$ and the latent variables $(\omega_1, \ldots, \omega_n)$ are used in a Gaussian scale mixture framework to represent a variety of common regression models for both binary and continuous data. In particular, we consider Gaussian linear regression~(see Section~\ref{sec:linear:regression}), linear regression with Laplace errors (see Section~\ref{sec:linear:laplace}), linear regression with Student $t$ errors (see Section~\ref{sec:linear:studentt}), and binary logistic regression~(see Section~\ref{sec:logistic}). The hyperparameters $(\tau^2 > 0)$ and $(\lambda_1,\ldots,\lambda_n)$ will be used to model different sparsity inducing prior distributions for the regression coefficients. The prior distributions considered in this paper are examples of common global--local shrinkage priors and include: the lasso (see Section~\ref{sec:prior:lasso}), ridge regression (see Section~\ref{sec:prior:ridge}), the horseshoe (see~Section~\ref{sec:prior:hs}), and the horseshoe$+$ estimator (see Section~\ref{sec:prior:hsplus}). Details of the software toolbox implementing these Bayesian penalized regression techniques are given in Section~\ref{sec:toolbox}.
 
%
%
\section{Bayesian penalized regression}
\label{sec:bayesian:penalised:regression}
Bayesian inference for a statistical model requires the posterior distributions for all model parameters. In the case of the penalized regression hierarchy (\ref{eqn:y})--(\ref{eqn:tau2}), analytical computation of the posterior distributions is intractable and we instead use MCMC techniques to obtain stochastic approximations to the corresponding posterior densities. Due to the particular choices of the prior distributions~(see Section~\ref{sec:priors}), all conditional posterior distributions are computable analytically, and sampling from these posterior densities can be done with the Gibbs sampler~\citep{GemanGeman84}.

We use the uniform prior distribution for the intercept parameter (\ref{eqn:beta0}) while the regression parameters are given a joint prior distribution specified by the Gaussian variance mixture (\ref{eqn:beta})--(\ref{eqn:tau2}). Given the aforementioned choice of prior distributions, the conditional posterior distributions for the intercept parameter $\beta_0$ and the regression parameters $\bm{\beta}$ are equivalent across the statistical models examined in this paper. Let
\begin{eqnarray}
	e_i &=& z_i - {\bf x}_i^\T \bm{\beta} - \beta_0, \quad (i=1,\ldots,n), \label{eqn:ei}\\
	\bm{\Omega}_n &=& \sigma^2 \, {\rm diag}(\omega_1^2, \ldots, \omega_n^2), \label{eqn:Omega} \\
	\bm{\Lambda}_p &=& \sigma^2 \tau^2 \, {\rm diag}(\lambda_1^2, \ldots, \lambda_p^2),
\end{eqnarray}
where $(e_1, \ldots, e_n)$ denote the model residuals in the case of linear regression and $\bm{\Omega}_n \in \mathbb{R}^{n \times n}$ and $\bm{\Lambda}_p \in \mathbb{R}^{p\times p}$ are diagonal matrices. In the case of linear regression models (see Section~\ref{sec:linear:regression}), the data $y_i$ is continuous and we set $z_i = y_i$, while in binary logistic regression (see Section~\ref{sec:logistic}) 
\begin{equation}
	z_i = \omega_i^2 \left(y_i - \frac{1}{2}\right)
\end{equation}
for all $i=1,\ldots,n$. In both cases, the conditional posterior distribution for the intercept parameter $\beta_0$ is the Gaussian distribution ${\rm N}(\tilde{\mu}, \tilde{\sigma}^2)$, where
\begin{equation}
	\tilde{\mu} =  \left(\sum_{i=1}^n \frac{z_i - {\bf x}_i^\T \bm{\beta}}{\omega^2_i} \right) \left(  \sum_{i=1}^n \frac{1}{\omega^2_i} \right)^{-1}, \quad
	\tilde{\sigma}^2 = \sigma^2 \left( \sum_{i=1}^n \frac{1}{\omega^2_i} \right)^{-1}.
\end{equation}
The conditional posterior density for the regression parameters $\bm{\beta}$ is the $p$-variate Gaussian distribution ${\rm N}_p(\tilde{\bm{\mu}}, {\bf A}_p^{-1})$, for which
\begin{equation}
	\tilde{\bm{\mu}} =  {\bf A}_p^{-1}{\bf X}^\T \bm{\Omega}_n^{-1} ({\bf z} - \beta_0 {\bf 1}_n), \quad
	{\bf A}_p = \left({\bf X}^\T \bm{\Omega}_n^{-1} {\bf X} + \bm{\Lambda}_p^{-1}\right), 
\end{equation}
where ${\bf 1}_n$ is an $n$-dimensional vector of ones and ${\bf A}_p \in \mathbb{R}^{p \times p}$ is a symmetric positive definite precision matrix. A detailed derivation of this conditional distribution is available in the seminal paper by~\cite{LindleySmith72}. 

In terms of computational efficiency, a major bottleneck of the proposed Gibbs algorithm is the sampling of the regression coefficients from the $p$-variate Gaussian distribution ${\rm N}_p(\tilde{\bm{\mu}}, {\bf A}_p^{-1})$ when the number of predictors $p$ is large. Direct computation of the matrix inverse ${\bf A}_p^{-1}$ is not recommended because it exhibits poor numerical accuracy and is computationally expensive. Instead, our implementation uses two algorithms for sampling the regression coefficients, where the choice of algorithm depends on the sample size $n$ and the number of regressors $p$. Specifically, we use the algorithm in ~\cite{Rue01} when the ratio $(p/n < 2)$ and ~\cite{Bhattacharya2016} otherwise. 

Rue's algorithm uses Cholesky factorization of the conditional posterior variance matrix ${\bf A}_p^{-1}$ and has computational complexity of the order $O(p^3)$. The algorithm is efficient as long as $p$ is not too large compared with $n$. In the case where $p$ is much larger than $n$, our sampler uses the algorithm in~\cite{Bhattacharya2016} which has computational complexity $O(n^2p)$, which is linear in $p$. \cite{Bhattacharya2016} show that their algorithm is orders of magnitude more efficient than Rue's algorithm when $(p/n > 2)$. Subsequently, sampling of the regression coefficients in the proposed toolbox is significantly faster than alternative implementations (e.g., {\tt Stan}) and represents the current state-of-the-art in terms of speed and numerical accuracy.

In the following sections, we describe the Gaussian scale mixture framework for the data (\ref{eqn:y})--(\ref{eqn:omega}) for linear (Section~\ref{sec:linear:regression}) and logistic (\ref{sec:logistic}) regression models.
\subsection{Linear regression}
\label{sec:linear:regression}
The Bayesian penalized regression hierarchy (\ref{eqn:y})--(\ref{eqn:tau2}) is easily adapted to the setting of Bayesian linear regression models with Gaussian noise. In the case of contaminated data or data with outliers, the Gaussian noise model is no longer appropriate and error distributions with heavier tails are required. The decision to model the data generating distribution as the Gaussian scale mixture (\ref{eqn:y})--(\ref{eqn:omega}) with mixing parameters $(\omega^2_1, \ldots, \omega^2_n)$ allows for a wide range of non-Gaussian, heavy-tailed noise models to be easily incorporated into the main hierarchy. In particular, we implement Laplace and Student $t$ noise models, which correspond to Gaussian scale mixtures with exponential and inverse gamma mixing densities, respectively.

The model for the noise is obtained from the prior distributions for the scale parameter $\sigma^2$ and the latent variables $(\omega^2_1, \ldots, \omega^2_n)$. Recall that the data ${\bf y}$ is continuous in the case of linear regression and we set $(z_i = y_i)$ in (\ref{eqn:y}) and (\ref{eqn:ei}). The treatment for binary data in the case of logistic regression is described in Section~\ref{sec:logistic}. For all noise models considered, the scale parameter $\sigma^2$ is given the scale invariant prior distribution $\pi(\sigma^2) \propto 1 / \sigma^2$. The conditional posterior distribution for $\sigma^2$ is the inverse gamma distribution ${\rm IG}(\tilde{\alpha}, \tilde{\beta})$, where
\begin{equation}
	\tilde{\alpha} = \frac{n + p}{2}, \quad \tilde{\beta} = \frac{1}{2} \left( \sum_{i=1}^n \frac{e_i^2}{\omega_i^2} + \sum_{j=1}^p \frac{\beta_j^2}{\tau^2 \lambda_j^2} \right). 
\end{equation}
Sampling of the latent variables $(\omega^2_1, \ldots, \omega^2_n)$ for Gaussian, Laplace and Student $t$ noise models is described in the following sections.

\subsubsection{Gaussian errors}
\label{sec:linear:gaussian}
In the case of Gaussian errors, the data is assumed to be generated by a single Gaussian distribution, not a Gaussian scale mixture distribution. The latent variables $(\omega^2_1, \ldots, \omega^2_n)$ are therefore set to $\omega^2_i = 1$ for all $i = (1, \ldots, n)$ requiring no sampling, which implies that $\bm{\Omega}_n = \sigma^2 {\bf I}_n$ in (\ref{eqn:Omega}).
\subsubsection{Laplace errors}
\label{sec:linear:laplace}
The Laplace distribution has heavier tails than the Gaussian distribution and is commonly used to model contaminated data and data with outliers. An important advantage of the Laplace distribution, over other heavy-tailed distributions such as the Student $t$, is that all of its central moments are finite. We represent the Laplace distribution as a Gaussian variance mixture distribution where the mixing density (\ref{eqn:gaussian:mixtures}) is
\begin{equation}
	\omega_i^{2} \sim {\rm Exp}(1), 
\end{equation}
which is an exponential distribution with a mean of one. This particular choice of the mixing distribution ensures that the residuals $(e_1, \ldots, e_n)$ follow a Laplace distribution and that the scale parameter $\sigma^2$ (see (\ref{eqn:sigma2})) is equal to the variance of the residuals, as in linear regression with Gaussian noise. The conditional posterior distribution of the latent variables $1/\omega_i^2$ is the inverse Gaussian distribution ${\rm IGauss}(\tilde{\mu}_i, \tilde{\lambda})$, where 
\begin{equation}
\tilde{\mu}_i = \left( \frac{2 \sigma^2}{e_i^2} \right)^{\frac{1}{2}}, \quad
\tilde{\lambda} = 2.
\end{equation}
for all $i = (1,\ldots,n)$.

\subsubsection{Student $t$ errors}
\label{sec:linear:studentt}
An alternative to the Laplace distribution commonly used in linear regression with contaminated data is the Student $t$ distribution. The Student $t$ distribution has heavier tails than the Gaussian distribution and is parameterized by a location, a scale and a degrees of freedom parameter $(\delta > 0$) that determines the heaviness of the tails. When the degrees of freedom parameter $(\delta = 1)$, the Student $t$ distribution reduces to the Cauchy distribution, which has heavier tails than both the Gaussian and Laplace distributions. In addition, the Student $t$ distribution has heavier tails than the Laplace distribution for all $(\delta \leq 5)$, but, unlike the Laplace distribution, the Student $t$ distribution has infinite variance when $(\delta \leq 2)$.

From~(\ref{eqn:studentt}), the Student $t$ distribution may be written as a Gaussian variance mixture distribution where the mixing density is the inverse gamma distribution
\begin{equation}
	\omega_i^{2} \sim {\rm IG}\left(\frac{\delta}{2}, \frac{\delta}{2}\right), \quad (i = 1,\ldots, n).
\end{equation}
The choice of this inverse gamma distribution as the prior distribution of the latent variables $(\omega_1, \ldots, \omega_n)$ ensures that the residuals follow a Student $t$ distribution with $\delta$ degrees of freedom and that the variance of the residuals $(e_1, \ldots, e_n)$ is related to the scale parameter $\sigma^2$ (see (\ref{eqn:sigma2})) by
\begin{equation}
	{\rm Var}(e_i) = \sigma^2 \left( \frac{\delta}{\delta - 2} \right), \quad (i = 1,\ldots, n)
\end{equation}
which is finite for all $(\delta > 2)$. Given the scale parameter $\sigma^2$ and the residuals $(e_1, \ldots, e_n)$, the conditional posterior distribution for the latent variables $\omega_i^2$ is the inverse gamma distribution ${\rm IG}(\tilde{\alpha}, \tilde{\beta}_i)$ where
\begin{equation}
\tilde{\alpha} = \frac{\delta + 1}{2}, \quad
\tilde{\beta}_i = \frac{1}{2} \left( \frac{e_i^2}{\sigma^2} + \delta \right).
\end{equation}
\subsection{Binary logistic regression}
\label{sec:logistic}
In the case of binary data $y_i \in  \{0, 1\}$ $(i = 1, \ldots, n)$, the relationship between the predictor variables and the outcome variable is represented using binary logistic regression models. In particular, we assume that 
\begin{equation}
	p(y_i = 1 | {\bf x}_i, \beta_0, \bm{\beta}) = \frac{1}{1 + \exp\left(-(\beta_0 + {\bf x}_i^\T \bm{\beta})\right)}.
\end{equation}
Direct sampling from the posterior distribution of the regression parameters in this binary logistic regression model is difficult due to the mathematical form of the logistic function. Recently, indirect sampling algorithms based on auxiliary (or latent) variables for logistic regression have been proposed by \cite{HolmesHeld06}, \cite{FruehwirthSchnatter07}, \cite{GramacyPolson12} and \cite{PolsonEtAl13}. Of these, the algorithm by \cite{PolsonEtAl13} is the current state-of-the-art in terms of computational and sampling efficiency as well as ease of implementation. Importantly, \cite{PolsonEtAl13} represent the logistic function as a Gaussian variance mixture distribution with a P\'{o}lya-gamma mixing density, which is easily integrated into the hierarchy (\ref{eqn:y})--(\ref{eqn:tau2}).

Implementation of Bayesian logistic regression with the P\'{o}lya-gamma representation requires the latent variables $(\omega_1,\ldots,\omega_n)$ and the scale parameter $\sigma^2$ to be sampled appropriately. Unlike in the case of linear regression, the scale parameter $\sigma^2$ will not require sampling and is fixed at $\sigma^2 = 1$. Following~\cite{PolsonEtAl13}, the latent variables $(\omega_1,\ldots,\omega_n)$ are given a P\'{o}lya-gamma prior distribution
\begin{equation}
	\omega_i^2 \sim {\rm PG}(0, 1).
\end{equation}
The conditional posterior distribution of the latent variables $1/\omega_i^2$ is the P\'{o}lya-gamma distribution ${\rm PG}(1, \tilde{c}_i)$, where
\begin{equation}
	\tilde{c}_i = \beta_0 + {\bf x}_i^\T \bm{\beta}, \quad (i=1,\ldots,n).
\end{equation}
An efficient algorithm for sampling from the P\'{o}lya-gamma distribution was recently proposed by \cite{WindleEtAl14} and is used in this software toolbox. The {\tt MATLAB} version of our toolbox uses a {\tt C++} implementation of the P\'{o}lya-gamma sampler, which is a direct conversion of the {\tt R} and {\tt C} code provided by \cite{WindleEtAl14}. The {\tt R} version of the toolbox currently depends on the package {\tt BayesLogit} which implements the same algorithm for sampling from P\'{o}lya-gamma random variables. 

\subsection{Prior distributions}
\label{sec:priors}
All prior distributions for the regression coefficients (\ref{eqn:beta})--(\ref{eqn:tau2}) considered in this paper are exchangeable Gaussian variance mixture distributions. Here, the latent variables $\tau^2$ and $(\lambda_1, \ldots, \lambda_p)$ determine the type of sparsity that is enforced on the regression coefficients $\bm{\beta}$. The hyperparameter $\tau^2$ corresponds to the global variance parameter that controls the amount of overall shrinkage of the coefficients. \cite{PolsonScott10} recommend that the prior distribution for $\tau^2$ should have substantial prior mass in the neighbourhood of zero to shrink the regression parameters and suppress noise.  

In the absence of expert prior knowledge, the prior distribution for the global shrinkage parameter $\tau$ is chosen to be 
\begin{equation}
	\tau \sim {\rm C}^{+}(0, 1), \label{eqn:tau:hc}
\end{equation}
where ${\rm C}^{+}(0, 1)$ is the half-Cauchy distribution with a mean of zero and a scale parameter of one. \cite{PolsonScott10} recommend the half-Cauchy distribution for $\tau$ as a sensible default, which also agrees with the findings of \cite{Gelman06}. Interestingly, the half-Cauchy distribution can be written as a mixture of inverse gamma distributions~\citep{MakalicSchmidt16a} so that (\ref{eqn:tau:hc}) is equivalent to:
\begin{equation}
	\tau^2 | \xi \sim {\rm IG}(1/2, 1/\xi), \quad
	\xi \sim {\rm IG}(1/2, 1), \label{eqn:tau:mixture}
\end{equation}
where $\xi > 0$ is a mixing parameter and ${\rm IG}(\tilde{\alpha},\tilde{\beta})$ is the inverse gamma distribution with shape parameter $\tilde{\alpha}$ and scale parameter $\tilde{\beta}$. The advantage of the mixture representation is that it allows sampling from the conditional posterior distributions of the latent variables $\tau$ and $(\lambda_1, \ldots, \lambda_p)$ with the Gibbs sampler. 

Assuming the latent variable representation (\ref{eqn:tau:mixture}), the conditional posterior distribution of $\tau^2$ is the inverse gamma distribution ${\rm IG} (\tilde{\alpha}, \tilde{\beta})$, where
\begin{equation}
	\tilde{\alpha} = \frac{p + 1}{2}, \quad
	\tilde{\beta} = \frac{1}{\xi} + \frac{1}{2\sigma^2}\sum_{j=1}^p \frac{\beta_j^2}{\lambda_j^2}.
\end{equation}
Similarly, the conditional posterior distribution for the mixing parameter $\xi$ is the inverse gamma distribution ${\rm IG} (\tilde{\alpha}, \tilde{\beta})$, where 
\begin{equation}
	\tilde{\alpha} = 1, \quad
	\tilde{\beta} = 1 + \frac{1}{\tau^2}.
\end{equation}

The hyperparameters $(\lambda_1, \ldots, \lambda_p)$ correspond to the local variance (shrinkage) components that determine the type of shrinkage penalty applied to the regression coefficients. Following~\cite{PolsonScott12a}, the local variance hyperparameters should have a prior distribution with: (i)~a pole at zero to guarantee predictive efficiency in recovering the true sampling distribution of the regression parameters $\bm{\beta} \in \mathbb{R}^p$ and (ii)~polynomial tails to aggressively shrink noise variables while allowing large signals to remain unchanged. Sampling details for the Bayesian ridge, lasso, horseshoe and horseshoe$+$ estimators are considered in the following sections. Of these estimators, only the horseshoe and horseshoe$+$ estimators possess the two desirable shrinkage properties.

\subsubsection{Ridge regression}
\label{sec:prior:ridge}
Bayesian ridge regression requires only a single global shrinkage hyperparameter $\tau$ and does not include any local shrinkage hyperparameters. Consequently, we set $(\lambda_j = 1)$ for all $j=(1,\ldots,p)$, which implies that sampling of these hyperparameters is not required.
\subsubsection{Lasso regression}
\label{sec:prior:lasso}
The Bayesian lasso estimator~\citep{ParkCasella08,Hans09} requires using a Laplace prior distribution for the regression coefficients so that the mode of the resulting posterior density corresponds to the usual lasso estimator~\citep{Tibshirani96}. The Laplace distribution may be represented as a Gaussian variance mixture distribution where the mixing density is an exponential distribution. The hyperparameters $(\lambda_1,\ldots,\lambda_p)$ therefore follow the exponential distribution
\begin{equation}
\label{eqn:lambda2:lasso}
	\lambda_j^2 \sim {\rm Exp}(1), \quad (j=1,\ldots,p),
\end{equation}
with a mean of one. Following~\cite{ParkCasella08}, the conditional posterior distribution for the local shrinkage parameter $1 / \lambda_j^2$ is the inverse Gaussian distribution ${\rm IGauss}(\tilde{\mu}_j, \tilde{\lambda})$, where
\begin{equation}
	\tilde{\mu}_j = \left(\frac{2 \tau^2 \sigma^2}{\beta_j^2}\right)^{\frac{1}{2}}, \quad
	\tilde{\lambda} = 2.
\end{equation}
for all $j=1,\ldots,p$. The hierarchy (\ref{eqn:tau:mixture}) and (\ref{eqn:lambda2:lasso}) is slightly different to the original Bayesian lasso proposal by \cite{ParkCasella08} where the hyperparameters $(\lambda_1,\ldots,\lambda_p)$ are assigned a joint exponential prior distribution which depends on a further hyperparameter. Our Bayesian lasso hierarchy moves the global shrinkage parameter $\tau^2$ down the hierarchy and to the same level as the local shrinkage hyperparameters which allows us to express the Bayesian lasso as a global-local shrinkage estimator, such as the horseshoe and the horseshoe$+$. Another advantage of our formulation over the original Bayesian lasso is that our hierarchy alleviates the need to specify hyperparameters at the highest level of the hierarchy.

\subsubsection{Horseshoe regression}
\label{sec:prior:hs}
Unlike the lasso and ridge prior distributions, the horseshoe prior exhibits a pole at zero and polynomial tails, important properties that guarantee good performance in the big data domain~\citep{PolsonScott12a}. The prior distribution for the local shrinkage hyperaprameters $(\lambda_1,\ldots,\lambda_p)$ is the zero-mean half-Cauchy distribution
\begin{equation}
	\lambda_j \sim \mathcal{C}^{+}(0, 1),
\end{equation}
which can equivalently be written as
\begin{eqnarray}
	\lambda_j^2 | \nu_j &\sim& {\rm IG}(1/2, 1/\nu_j),\\
	\nu_j &\sim& {\rm IG}(1/2, 1).
\end{eqnarray}
The conditional posterior distribution for the local shrinkage parameter $(1 / \lambda_j^2)$ is the exponential distribution ${\rm Exp}(\tilde{\beta}_j)$, where
\begin{equation}
	\tilde{\beta}_j =  \frac{1}{\nu_j} + \frac{\beta_j^2}{2\tau^2\sigma^2}. 
\end{equation}
%
%
%
Similarly, the conditional posterior distributions for the hyperparameter $(1 / \nu_j)$ is the exponential distribution ${\rm Exp}(\tilde{\beta}_j)$, where
\begin{eqnarray}
	\tilde{\beta}_j =  1 + \frac{1}{\lambda^2_j}.
\end{eqnarray}
The conditional distributions for the hyperparameters $\lambda_j$ and $\nu_j$ are exponential distributions for which efficient sampling algorithms exist, even when the number of predictors $p$ is large.
%

\subsubsection{Horseshoe$+$ regression}
\label{sec:prior:hsplus}
The horseshoe$+$ estimator~\citep{BhadraEtAl15} is a natural extension of the horseshoe estimator to ultra-sparse problems. In contrast to the horseshoe estimator, the horseshoe$+$ estimator has a lower posterior mean squared error and faster posterior concentration rates in terms of the Kullback--Leibler divergence metric. As with the horseshoe estimator, the prior distribution for the local shrinkage hyperaprameters $(\lambda_1,\ldots,\lambda_p)$ is the zero-mean half-Cauchy distribution, where
\begin{eqnarray}
	\lambda_j &\sim& \mathcal{C}^{+}(0, \phi_j), \\
	\phi_j &\sim& \mathcal{C}^{+}(0, 1).
\end{eqnarray}
The key difference between the horseshoe and horseshoe$+$ estimators is that the horseshoe$+$ incorporates an extra level of hyperparameters $(\phi_1, \ldots, \phi_p)$, where each $\phi_j$ corresponds to the prior variance associated with the hyperparameter $\lambda_j$. Recalling the parameter expansion proposed in~\cite{MakalicSchmidt16a}, the horseshoe$+$ hierarchy is equivalent to:
\begin{eqnarray}
	\lambda_j^2 | \nu_j &\sim& {\rm IG}(1/2, 1/\nu_j),\\
	\nu_j | \phi_j^2 &\sim& {\rm IG}(1/2, 1/\phi_j^2), \\
	\phi_j^2 | \zeta_j  &\sim& {\rm IG}(1/2, 1/\zeta_j), \\
	\zeta_j &\sim& {\rm IG}(1/2, 1). 
\end{eqnarray}
This enables straightforward computation of the posterior conditional distributions for all hyperparameters. In particular, the conditional posterior distribution for the local shrinkage parameters $1 / \lambda_j^2$ is the exponential distribution ${\rm Exp}(\tilde{\beta}_j)$, where 
%
%
\begin{equation}
	\tilde{\beta}_j =  \frac{1}{\nu_j} + \frac{\beta_j^2}{2\tau^2\sigma^2},
\end{equation}
which is equivalent to the corresponding conditional posterior density in the horseshoe estimator. The conditional posterior densities for the other hyperparameters are:
\begin{eqnarray}
1 / \nu_j &\sim& {\rm Exp} \left(\frac{1}{\phi_j^2} + \frac{1}{\lambda^2_j}\right), \\
1 / \phi_j^2 &\sim& {\rm Exp} \left( \frac{1}{\nu_j} + \frac{1}{\zeta_j} \right), \\
1 / \zeta_j  &\sim& {\rm Exp} \left( 1 + \frac{1}{\phi^2_j} \right). 
\end{eqnarray}

\section{Software implementation}
\label{sec:toolbox}
We have implemented the proposed Bayesian penalized regression toolbox for the {\tt MATLAB} and {\tt R} programming environments. Within both platforms, the toolbox can be accessed via the function {\tt bayesreg}. Due to fundamental differences between {\tt R} and {\tt MATLAB}, the syntax of the {\tt bayesreg} function is slightly different, but all the {\tt bayesreg} command line options and parameters are equivalent across the two platforms. As such, we only discuss the MATLAB format of {\tt bayesreg} below, but examples of using {\tt bayesreg} for both {\tt MATLAB} and {\tt R} are given in Section~\ref{sec:toolbox:examples}. Where appropriate, any significant differences between the versions will be noted. The toolbox is available for download from the CRAN repository for {\tt R} packages (package name {\tt bayesreg}) and the MATLAB Central File Exchange (File ID \#60823).
\subsection{{MATLAB}/R toolbox}
\label{sec:toolbox:matlab}
The syntax for {\tt bayesreg} is:
\begin{footnotesize}
\begin{verbatim}
[beta, beta0, retval] = bayesreg(X, y, model, prior, ...); % MATLAB version
bayesreg <- function(formula, data, model='normal', prior='ridge', ...) # R version 
\end{verbatim}
\end{footnotesize}
where:
\begin{itemize}
\item {\tt X} is an $(n \times p)$ matrix of predictors that does not contain the constant
\item {\tt y} is the $(n \times 1)$ output vector that may be binary ($y \in \{0,1\}^n$) or continuous ($y \in \mathbb{R}^n$)
\item {\tt formula} and {\tt data} are standard {\tt R} mechanisms for regression
\item {\tt model} is the error model, which may be {\tt 'gaussian'} (see Section~\ref{sec:linear:gaussian}), {\tt 'laplace'} (see Section~\ref{sec:linear:laplace}), {\tt 't'} (see Section~\ref{sec:linear:studentt}), or {\tt 'binomial'} (see Section~\ref{sec:logistic})
\item {\tt prior} is the prior density for the regression coefficients, which can be set to {\tt 'ridge'} (see Section~\ref{sec:prior:hsplus}), {\tt 'lasso'} (see Section~\ref{sec:prior:lasso}), {\tt 'hs'} (see Section~\ref{sec:prior:hs}), or {\tt 'hs+'} (see Section~\ref{sec:prior:hsplus}). 
\end{itemize}
By default, {\tt bayesreg} generates $1,000$ samples from the posterior distribution using a burnin period of $1,000$ samples and a thinning level of $5$. The first four {\tt bayesreg} arguments (i.e., {\tt X, y, model} and  {\tt prior}) are mandatory in {\tt MATLAB}. The following optional arguments are supported:
\begin{itemize}
\item {\tt nsamples} -- number of samples to draw from the posterior distribution (default: $1,000$)
\item {\tt burnin} -- number of burnin samples (default: $1,000$)
\item {\tt thin} -- level of thinning (default: $5$)
\item {\tt display} -- whether summary statistics are printed (default: {\tt true})
\item {\tt displayor} -- display odds ratios instead of regression coefficients in the case of logistic regression? (default: {\tt false})
\item {\tt varnames} -- a cell array containing names of the predictor variables (default: {\tt 'v1', 'v2', etc.})
\item {\tt sortrank} -- display the predictors in the order of importance as determined by the variable rank? (default: {\tt false})
\item {\tt tdof} -- degrees of freedom for the $t$-distribution (default: $5$).
\end{itemize}

The {\tt MATLAB} version of {\tt bayesreg} prints out a table of summary statistics after the sampling is completed. The same summary statistics can be obtained in {\tt R} using the {\tt summary()} command on the object returned by {\tt bayesreg}. The summary statistics include the posterior mean, the posterior standard deviation and the 95\% credible interval for each regression coefficient. In addition, {\tt bayesreg} displays an estimate of the $t$-statistic (i.e., the posterior mean divided by the posterior standard deviation), and the rank and the effective sample size for each of the $p$ predictor variables. The rank statistic is estimated with the Bayesian feature ranking algorithm~\citep{MakalicSchmidt11c} and corresponds to the strength of the association between the variable and the target data ${\bf y}$. The estimated rank $r$ of a variable is an integer ($1 \leq r \leq p$) where lower ranks denote more important variables. That is, a variable with rank $(r = 1)$ is deemed to have the strongest association with the target, while the variable with rank $(r = p)$ is estimated to be least associated with ${\bf y}$. The effective sample size diagnostic from \cite{RobertCasella04} (pp. 499--500) and \cite{Geyer92} is expressed as a percentage and determines the sampling efficiency of the chains for each predictor. The function {\tt bayesreg} computes the effective sample size from the autocorrelation of the sampling chains. Lastly, {\tt bayesreg()} may display one or two asterisk (*) symbols next to the rank of each variable. The first (second) asterisk is printed when the 75\% credible interval (95\% credible interval) for the corresponding predictor does not include 0.

Upon completion, the MATLAB version of {\tt bayesreg} returns the following: 
\begin{itemize}
\item {\tt beta}, a matrix of posterior samples of the regression coefficients $\bm{\beta}$ of size $(p \times N)$, where $p$ is the number of predictors and $N$ is the number of posterior samples
\item {\tt beta0}, a vector of posterior samples for the intercept parameter of size $(1 \times N)$
\item {\tt retval}, an object of type {\tt struct} that contains posterior samples of the model hyperparameters and additional sampling statistics, such as the posterior mean of the regression coefficients.
\end{itemize}
The {\tt R} version returns an object of class {\tt bayesreg} which contains the posterior samples of all the parameters and hyperparameters as well as additional sampling statistics. Examples of using {\tt bayesreg} are presented in the following section.
\subsection{Examples}
\label{sec:toolbox:examples}
Below are some typical usage examples for {\tt bayesreg} under the {\tt MATLAB} and {\tt R} programming platforms. We begin by generating data from a linear regression model where we know the true parameter coefficients $\bm{\beta} \in \mathbb{R}^p$. In {\tt MATLAB}, the data can be generated with the following commands:

\begin{footnotesize}
\begin{verbatim}
>> clear;
>> n = 50;                                % Sample size
>> p = 10;                                % Number of predictors
>> rho = 0.5;                             % Covariance of X is AR(1) with homogenous variances 
>> S = toeplitz(rho.^(0:p-1));
>> X = mvnrnd(zeros(p,1), S, n);          % Generate predictor matrix
>> b = [5;3;3;1;1; zeros(p-5,1)];         % True regression coefficients
>> snr = 4;                               % Signal-to-noise ratio
>> mu = X*b;
>> s2 = var(mu) / snr;                    % Residual variance
>> y = mu + sqrt(s2)*randn(n,1);          % Generate data y
\end{verbatim}
\end{footnotesize}
In {\tt R}, the equivalent commands to generate test data are:
\begin{footnotesize}
\begin{verbatim}
> library(MASS)
> library(bayesreg)
> rm(list = ls())
> n <- 50                                 # Sample size
> p <- 10                                 # Number of predictors
> rho <- 0.5                              # Covariance of X is AR(1) with homogenous variances 
> S <- toeplitz(rho^(0:(p-1)))
> X <- mvrnorm(n = n, rep(0, p), S)       # Generate predictor matrix
> b <- as.vector(c(5,3,3,1,1,rep(0,p-5))) # True regression coefficients
> snr <- 4                                # Signal-to-noise ratio
> mu <- X%*%b
> s2 <- var(mu) / snr                     # Residual variance
> y <- mu + sqrt(s2)*rnorm(n)             # Generate data y
> df <- data.frame(X,y)
\end{verbatim}
\end{footnotesize}

We then use {\tt bayesreg} to fit a Bayesian penalized regression model with the horseshoe prior to the data. We generate $10,000$ samples from the posterior distribution, discard the first $10,000$ samples as burnin and use a thinning level of 10:

\begin{footnotesize}
\begin{verbatim}
% MATLAB
>> [beta, beta0, retval] = bayesreg(X,y,'gaussian','hs','nsamples',1e4,'burnin',1e4,'thin',10);	

# R
> rv <- bayesreg(y ~ ., data = df, model="gaussian", prior="hs+", nsamples=1e4, burnin=1e4, thin=10)
> rv.s <- summary(rv)
\end{verbatim}
\end{footnotesize}
In {\tt MATLAB}, the code automatically prints out a summary table upon completion of the sampling. To obtain the same summary statistics in {\tt R}, we use the {\tt summary} method which, in addition to printing a table, returns an object containing all the summary statistics.
\begin{footnotesize}
\begin{verbatim}
==========================================================================================
|                   Bayesian Penalised Regression Estimation ver. 1.70                   |
|                         (c) Enes Makalic, Daniel F Schmidt. 2016                       |
==========================================================================================
Bayesian linear horseshoe regression                            Number of obs   =       50
                                                                Number of vars  =       10
MCMC Samples   =  10000                                         Root MSE        =   3.8963
MCMC Burnin    =  10000                                         R-squared       =   0.7962
MCMC Thinning  =     10                                         DIC             =  -143.07

-------------+----------------------------------------------------------------------------
   Parameter |  mean(Coef)  std(Coef)    [95% Cred. Interval]      tStat    Rank       ESS
-------------+----------------------------------------------------------------------------
          v1 |     4.17213    0.76739      2.61210    5.62765      5.437       1 **   93.3
          v2 |     2.46840    0.99140      0.40715    4.38094      2.490       3 **   89.9
          v3 |     4.03917    0.86229      2.30542    5.71276      4.684       1 **   95.9
          v4 |     1.05788    0.68354     -0.08983    2.42359      1.548       5 *    85.9
          v5 |     0.86717    0.77785     -0.30471    2.54351      1.115       6 *    83.6
          v6 |     1.59415    0.84976     -0.00166    3.21233      1.876       4 *    88.9
          v7 |     0.06269    0.47543     -0.92641    1.13763      0.132       7      97.8
          v8 |    -0.09137    0.48461     -1.19403    0.90716     -0.189       7      92.1
          v9 |    -0.30255    0.59472     -1.68075    0.73269     -0.509       7      88.8
         v10 |    -0.16704    0.50695     -1.34197    0.78697     -0.330       7      92.0
       _cons |     0.34938    0.65599     -0.92804    1.64925      0.533       .         .
-------------+----------------------------------------------------------------------------
\end{verbatim}
\end{footnotesize}

The effective sample size for all $10$ predictors appears adequate, suggesting that the number of samples from the posterior distribution is sufficient for this problem. The first six predictors are ranked top by the Bayesian feature ranking algorithm, with the remaining four predictors given the lowest rank of seven. This ranking algorithm strongly suggests including predictors {\tt 'v1','v2','v3'} in the final model, as indicated by the two asterisk symbols next to the corresponding ranks. It may also be of interest to examine the sampling chains for the parameters and hyperparameters visually using the {\tt MATLAB} {\tt plot()} command:

\begin{footnotesize}
\begin{verbatim}
>> plot(beta0); grid; xlabel('b0');
>> plot(beta(3,:)); grid; xlabel('b3');
>> plot(retval.sigma2); grid; xlabel('\sigma^2');
>> boxplot(beta'); grid;
\end{verbatim}
\end{footnotesize}

The sampling chains for $\beta_0, \beta_3$, and $\sigma^2$ appear to have converged to the target posterior distribution. In {\tt MATLAB}, to make predictions using the model previously fitted with {\tt bayesreg}, we create a new variable {\tt yhat} and generate the predictions using the posterior mean estimates of the regression coefficients:

\begin{footnotesize}
\begin{verbatim}
>> muhat = retval.muB0 + X*retval.muB; % or, ...
>> muhat = mean(beta0) + X*mean(beta,2);
>> mean( (mu-muhat).^2 )
>> sqrt( mean( (mu-muhat).^2 ) )
\end{verbatim}
\end{footnotesize}

The last two commands compute the mean squared error and the root mean squared error for the predictions {\tt yhat}. Equivalently, to make predictions in {\tt R}, we use the {\tt predict} method as follows:
\begin{footnotesize}
\begin{verbatim}
muhat <- predict(rv, df)
mean( (mu-muhat)^2 )
sqrt( mean( (mu-muhat)^2 ) )
\end{verbatim}
\end{footnotesize}
As a side note, the {\tt predict} method for {\tt bayesreg} objects can also be used to generate conditional probabilities and best guesses at class assignment in the case of logistic regression.

It is possible to use {\tt bayesreg} to analyze regression models when the number of predictors is large. For example, the code below generates a predictor matrix ${\bf X}$ with $n=50$ samples and $p=50,000$ variables. We then generate the data ${\bf y}$ with Laplace noise and use {\tt bayesreg} to fit a Bayesian regression model with the horseshoe$+$ prior for the regression coefficients:

\begin{footnotesize}
\begin{verbatim}
>> clear;
>> n = 50;                            % Sample size
>> p = 5e4;                           % Number of predictors
>> X = randn(n,p);                    % Generate predictor matrix
>> btrue = [5;5;1;1;1; zeros(p-5,1)]; % True regression coefficients
>> snr = 8;                           % Signal-to-noise ratio
>> s2 = var(X*btrue) / snr;           % Residual variance
>> b = sqrt(s2/2);
>> y = X*btrue + lplrnd(0,b,n,1);     % Generate data y with Laplace noise
>> tic, ...                           % Run and time bayesreg
>> [beta, beta0, retval] = bayesreg(X,y,'laplace','hs+',...
>>     'nsamples',1e3,'burnin',1e3,'thin',1,'display',false); 
>> toc
\end{verbatim}
\end{footnotesize}

This code takes approximately 100 seconds to generate 2,000 posterior samples on a standard laptop computer (Intel Core i7 6600U CPU with 16 GB of RAM), which is orders of magnitude faster than performing the equivalent sampling using the {\tt R} package {\tt monomvn}~\citep{MakalicSchmidt16a} or a general purpose program such as {\tt Stan}. In the next section, we demonstrate how {\tt bayesreg} can be used to analyze a real data classification problem with Bayesian logistic regression.

\subsubsection{Diabetes data}
To demonstrate {\tt bayesreg} for Bayesian logistic regression, we use the Pima Indians diabetes data set, which was provided to the UCI Machine Learning Repository~\citep{Lichman13} by the National Institute of Diabetes and Digestive and Kidney Diseases. The data has ($n=768$) observations, collected from 21-year-old female patients of Pima Indian heritage, and $(p=8)$ predictor variables:
\begin{enumerate}
\item {\tt PREG} -- number of times pregnant
\item {\tt PLAS} -- plasma glucose concentration at 2 hours in an oral glucose tolerance test
\item {\tt BP} -- diastolic blood pressure (mm Hg)
\item {\tt SKIN} -- triceps skin fold thickness (mm)
\item {\tt INS} -- 2-hour serum insulin (mu U/ml)
\item {\tt BMI} -- body mass index (weight in kg/(height in m)$^2$)
\item {\tt PED} -- diabetes pedigree function
\item {\tt AGE} -- age (years).
\end{enumerate}

We use this data to train a logistic regression model to predict the outcome of diabetes. The {\tt MATLAB} code to do this with {\tt bayesreg} is:

\begin{footnotesize}
\begin{verbatim}
>> clear;
>> load data/pima.mat
>> [beta, beta0, retval] = bayesreg(X,y,'binomial','lasso', 'displayor', true, ...
         'nsamples',1e4,'burnin',1e4,'thin',5,'varnames',varnames);

==========================================================================================
|                   Bayesian Penalised Regression Estimation ver. 1.70                   |
|                         (c) Enes Makalic, Daniel F Schmidt. 2016                       |
==========================================================================================
Bayesian logistic lasso regression                              Number of obs   =      768
                                                                Number of vars  =        8
MCMC Samples   =  10000                                         Log. Likelihood =  -361.96
MCMC Burnin    =  10000                                         Pseudo R2       =   0.2713
MCMC Thinning  =      5                                         DIC             =  -370.59

-------------+----------------------------------------------------------------------------
   Parameter |  median(OR)    std(OR)    [95% Cred. Interval]      tStat    Rank       ESS
-------------+----------------------------------------------------------------------------
        PREG |     1.12457    0.03598      1.05662    1.19764      3.719       3 **   95.2
        PLAS |     1.03485    0.00377      1.02779    1.04257      9.374       1 **   92.5
          BP |     0.98932    0.00503      0.97956    0.99928     -2.099       5 **   97.6
        SKIN |     0.99994    0.00613      0.98815    1.01217     -0.011       7      90.8
         INS |     0.99908    0.00085      0.99733    1.00066     -1.083       7 *   100.0
         BMI |     1.08967    0.01631      1.05921    1.12316      5.788       2 **   96.0
         PED |     2.35554    0.74972      1.34925    4.28815      2.885       4 **   93.6
         AGE |     1.01364    0.00897      0.99694    1.03210      1.522       6 *    94.4
       _cons |     0.00026    0.00024      0.00006    0.00101    -11.561       .         .
-------------+----------------------------------------------------------------------------
\end{verbatim}
\end{footnotesize}

\begin{figure*}[htbp]
\begin{center}
\subfigure[Regression coefficients]{
   \includegraphics[width=6.5cm]{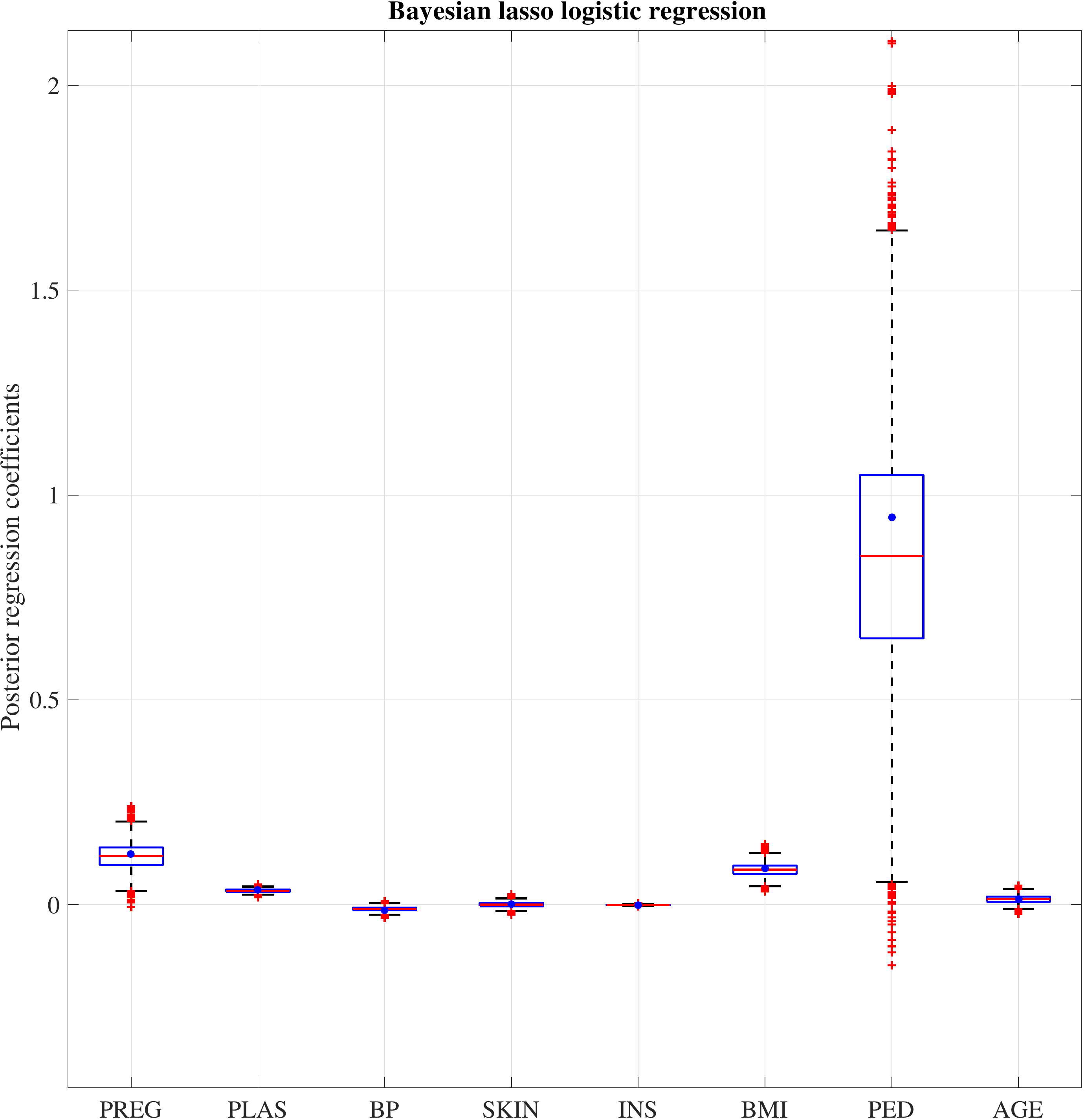}
}%
\hspace{1cm}
\subfigure[Probability density functions]{
   \includegraphics[width=6.5cm]{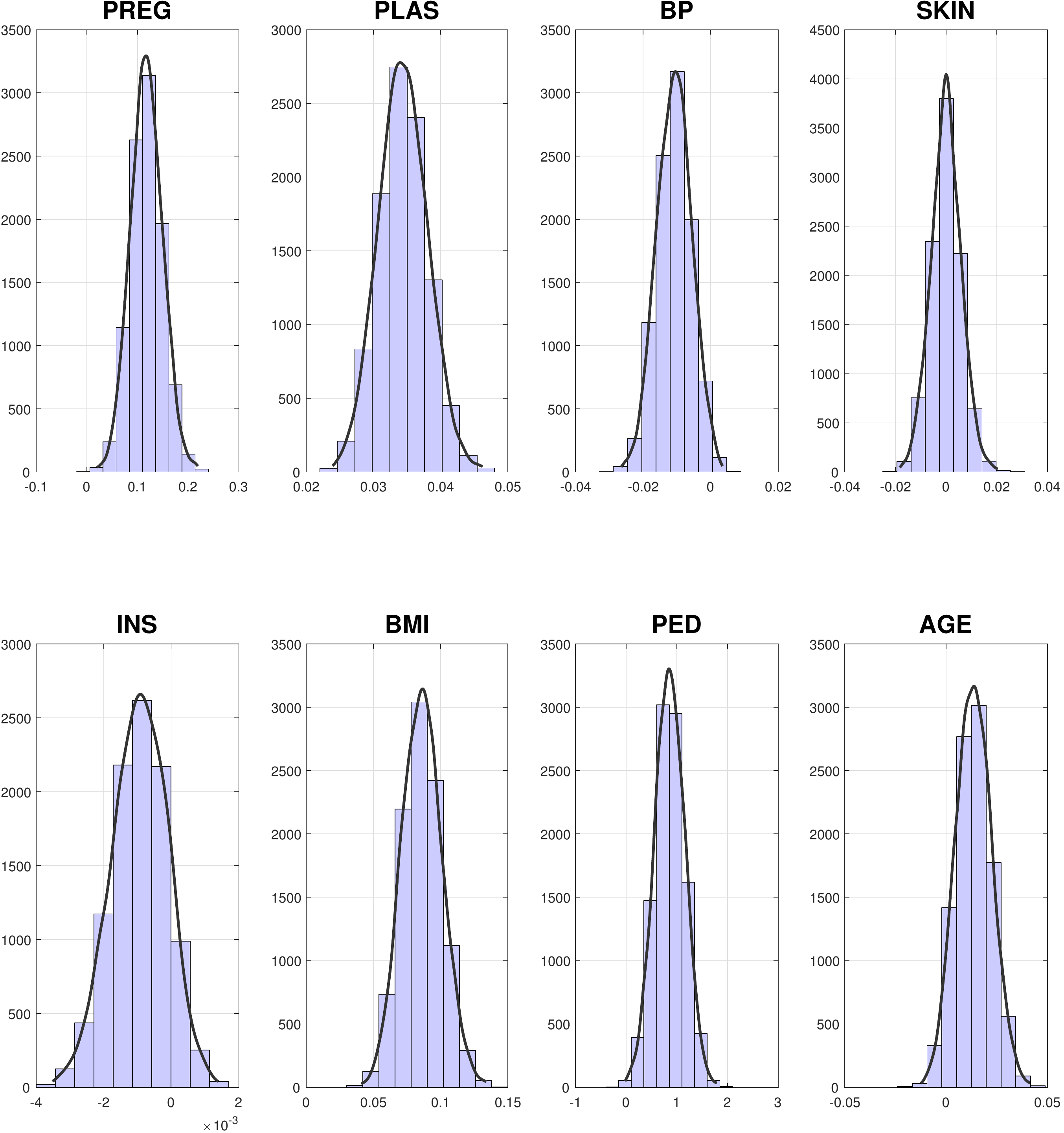}
}
\end{center}
\caption{Bayesian lasso logistic regression analysis of the Pima Indian diabetes data set. A boxplot of the posterior samples of the regression coefficients with the corresponding maximum likelihood estimator (left) and the estimated conditional posterior probability density functions (right).\label{fig:bayeslasso:pima}}
\end{figure*}

Here, we used the option {\tt displayor} to show posterior median odds ratios instead of the posterior mean regression coefficients, and we added the option {\tt varnames} to tell {\tt bayesreg} the names of the predictors. The odds ratios were estimated with respect to a unit change in the predictor variable. Figure~\ref{fig:bayeslasso:pima} shows a boxplot of the posterior samples of the regression coefficients and the estimated conditional posterior probability density functions. In this example, the Bayesian feature ranking algorithm combined with Bayesian logistic lasso regression ranks plasma glucose level as the most important variable, followed by BMI, number of pregnancies, family history of diabetes, and diastolic blood pressure. The three variables that were deemed not important were age, serum insulin level, and triceps skin fold thickness. 

\section{Discussion}
\label{sec:discussion}
The {\tt bayesreg} toolbox is the only toolbox for Bayesian penalized linear and logistic regression with continuous shrinkage priors that can readily be used for high dimensional problems. It implements linear regression with Gaussian and heavy-tailed error models, as well as logistic regression with the state-of-the-art P\'{o}lya-gamma data augmentation strategy. The toolbox supports four different types of continuous shrinkage priors allowing dense models (ridge regression) as well as varying levels of sparsity; e.g., Bayesian lasso for sparse models and horseshoe and horseshoe$+$ for ultra-sparse data. All sampling code is implemented using the Gibbs sampler which could potentially be combined with, for example, Chib's algorithm~\citep{Chib95} for computation of the marginal data likelihood. The toolbox also incorporates a simple metric for ranking of predictors via the $t$ statistic or the Bayesian feature ranking algorithm~\citep{MakalicSchmidt11c}. The {\tt bayesreg} toolbox, complete with full source code, is available for the {\tt MATLAB} (MATLAB Central File Exchange, File ID \#60823) and {\tt R} (CRAN, package name {\tt bayesreg}) numerical computing environments. We are also developing a version of {\tt bayesreg} for the statistical software package {\tt Stata}. 

To date, there are 21 packages on MATLAB Central File Exchange containing the search terms Bayesian regression. Of those, only two packages are in the same domain as {\tt bayesreg}; one of those packages implements Bayesian lasso linear regression while the other package is on variational Bayes for linear regression. The MATLAB function implementing Bayesian lasso linear regression uses direct matrix inversion to sample the regression coefficients $\bm{\beta} \in \mathbb{R}^p$ which is slow, numerically unstable and not suitable for data sets containing more than approximately $100$ predictors. There exist no implementations of Bayesian linear regression with heavy tailed error models, Bayesian logistic regression or the horseshoe and horseshoe$+$ estimators on MATLAB Central File Exchange.

The CRAN repository for {\tt R} packages contains several implementations of Bayesian shrinkage regression including five packages related to the current toolbox. Of those, three packages implement horseshoe regression with indirect posterior sampling algorithms based on the slice sampler~\citep{GramacyPantaleo09,HahnEtAl16,Bhattacharya2016}. The slice sampler by \cite{GramacyPantaleo09} is implemented in the {\tt R} package {\tt monomvn} and is significantly slower than our toolbox. For example, our {\tt bayesreg} implementation is approximately 40 times faster than {\tt monomvn} when tested with data where $n=$1,000 and $p=$1,000~\citep{MakalicSchmidt16a} while showing similar, if not better, rates of posterior convergence. 

\cite{HahnEtAl16} have proposed an elliptic slice sampler~\citep{MurrayEtAl10} for Bayesian linear regression that appears to offer some computational advantages over the Gibbs our sampling approach when applied to the horseshoe estimator. However, unlike our implementation, the elliptical slice sampler can only be used when the sample size is greater than the number of predictors and the design matrix is of full rank. Additionally, the elliptical slice sampler is significantly less flexible than {\tt bayesreg} and cannot easily be extended to handle grouped variables. In contrast, extension of our latent variable approach to handle multi-level groupings of variables (e.g., genetic markers grouped into genes, and genes grouped into pathways) is straightforward~\citep{XuElAl16}. An implementation of this elliptical slice sampler is available in the {\tt R} package {\tt fastHorseshoe}.

The package {\tt horseshoe}~\cite{Bhattacharya2016} also uses the slice sampler to implement Bayesian linear regression for the horseshoe estimator. The methodology for sampling of the regression coefficients is efficient and similar to that employed in {\tt bayesreg}. However, {\tt horseshoe} only implements Gaussian linear regression and does not allow for alternative prior distributions and error models.

In terms of features, the closest implementation of Bayesian shrinkage regression to our toolbox is the {\tt R} package {\tt rstanarm}. This package provides an R interface to the Stan C++ library for Bayesian estimation and features Bayesian shrinkage regression for continuous, binary and count data. The implementation is based on Hamiltonian Monte Carlo combined with the `No-U-Turn Sampler' (NUTS) sampler~\citep{HoffmanGelman14}. The package provides an interface to Stan implementations of recent shrinkage priors, including the horseshoe, the horseshoe$+$, the Dirichlet--Laplace~\citep{BhattacharyaEtAl15} and the R2--D2~\citep{ZhangEtAl16} estimator. Unsurprisingly, due to the general nature of Stan, Bayesian shrinkage regression with these priors within Stan is significantly slower than a specialized Gibbs sampler. For example, {\tt rstanarm} takes approximately 40s to obtain $2,000$ samples for Bayesian horseshoe Gaussian regression using a data set comprising $n = 442$ observations and $p = 10$ predictors. The equivalent operation takes approximately $0.15$s using the MATLAB version of {\tt bayesreg}. Furthermore, the NUTS sampler used within Stan appears to sometimes produce divergent MCMC transitions for the horseshoe and horseshoe$+$ estimators~\citep{PiironenVehtari06}.

\subsection{Future work}
Features planned for future versions of {\tt bayesreg} include:
\begin{itemize}
\item additional prior densities for the regression coefficients, such as the Dirichlet--Laplace~\citep{BhattacharyaEtAl15} and the R2--D2~\citep{ZhangEtAl16} estimators
\item negative binomial regression using the Gaussian scale mixture decomposition of~\cite{PolsonEtAl13}
\item longitudinal (Gaussian) linear regression with the variance--covariance matrix prior distribution proposed by~\cite{HuangWand13}
\item Bayesian autoregressive noise models~\citep{SchmidtMakalic13}
\item grouping of variables to allow better support of categorical (factor) data
\item block sampling of regression coefficients for ultra-high-dimensional data sets.
\end{itemize}


\bibliography{bibliography}

\end{document}